\documentclass[12pt]{article}
\usepackage{amsmath,amssymb,amscd,cite}
\newtheorem{thm}{Theorem}[section]
\newtheorem{prop}[thm]{Proposition}
\newtheorem{lem}[thm]{Lemma}
\newtheorem{cor}[thm]{Corollary}

\newtheorem{defi}[thm]{Definition}
\newcommand{\pf}{{\bf Proof. \ }}
\newcommand{\qed}{\hfill $\blacksquare$ \\}

\font\msbm=msbm10 at 12pt
\newcommand{\Z}{\mbox{\msbm Z}}

\newcommand{\F}{\mbox{\msbm F}}
\newtheorem{rem}[thm]{Remark}
\newtheorem{ex}[thm]{Example}

\date{}

\begin{document}

\title{On Codes over $\mathbb{F}_{q}+v\mathbb{F}_{q}+v^{2}\mathbb{F}_{q}$}
\author{A. Melakhessou, K. Guenda, T. A. Gulliver, M. Shi and P. Sol\'e
\thanks{A. Melakhessou is with Department of Mathematics, University of Batna 2, Algeria,
K. Guenda is with the Faculty of Mathematics USTHB, Algeria,
T. A. Gulliver is with the Department of Electrical and Computer Engineering, University of Victoria, Canada,
M. Shi is with the School of Mathematical Sciences, Anhui University, P.R. China, and
P. Sol\'e is with the Department of Mathematics,
Universit\'{e} de Vincennes - Paris 8,
France.}}
\maketitle
\begin{abstract}
In this paper we investigate linear codes with complementary dual (LCD) codes and formally self-dual codes over the ring $R=\F_{q}+v\F_{q}+v^{2}\F_{q}$, where $v^{3}=v$, for $q$ odd.
We give conditions on the existence of LCD codes and present construction of formally self-dual codes over $R$.
Further, we give bounds on the minimum distance of LCD codes over $\F_q$ and extend these to codes over $R$.
\end{abstract}

\section{Introduction}

Codes over finite rings have been known for several decades, but
interest in these codes increased substantially after the discovery that good non-linear binary
codes can be constructed from codes over $\Z_{4}$.
Recently, codes over finite non chain rings have been considered.
Codes over the ring $R=\F_{q}+v\F_{q}+v^{2}\F_{q}$, $v^3=v$ were introduced by Gao et al. \cite{gao},
and Shi et al. \cite{sol} studied their properties and gave several families of codes over this ring.

Linear codes with complementary dual (LCD) codes over finite fields were first studied by Massey \cite{massey} and more
recently by Carlet and Guilley \cite{carlet} and Dougherty et al. \cite{soleLCD}.
LCD codes can be used to protect information against side channel attacks \cite{carlet}. When considered over $\F_q$ these codes have the advantage to be easily decoded \cite{massey}. This property also applies to LCD codes over $R$
because this ring can be seen as the direct product $\F_q \times  \F_q \times  \F_q $.
Further, LCD codes can be used to obtain optimal entanglement-assisted quantum codes \cite{GJG}.

Formally self-dual codes are an important class of codes because
they have weight enumerators that are invariant under the MacWilliams transform,
and can have better parameters than self-dual codes \cite{Batoul}.
In this paper, LCD and formally self-dual codes are considered over the ring $R=\F_{q}+v\F_{q}+v^{2}\F_{q}$, where $v^{3}=v$ and $q$ is an odd prime power.
We give conditions on the existence of LCD codes over this ring.
Further, several constructions of LCD codes are given, in particular from weighing matrices.
Constructions of formally self-dual codes over $R$ are also presented.
In addition, bounds on the minimum distance of LCD codes over the finite field $\F_q$ are given and extended to codes over $R$.

The remainder of this paper is organized as follows.
Section 2 provides some preliminary results regarding the finite ring $R=\F_{q}+v\F_{q}+v^{2}\F_{q}$.
The Lee weights of the elements of $R$ are defined, and a Gray map is introduced.
This map leads to some useful results on linear codes over $R$.
We investigate the relationship between the dual and Gray image of codes.
Section 3 considers LCD codes over $R$.
Necessary and sufficient conditions on the existence of LCD codes over $R$ are given,
and LCD codes are constructed from weighting matrices.
Tables of LCD codes up to length $40$ are given which are obtained
from skew matrices and conferences matrices over the finite field
$\F_p$ where $p$ is a prime number such that $3< p \le 23$.
In Section 4, we present three constructions of formally self-dual codes over $R$.
Further, LCD codes are constructed which are also formally self-dual.
In Section 5 we give bounds on the minimum distance of LCD codes over $\F_q$,
and these bounds are extended to free LCD codes over $R$.

\section{Preliminaries}

In this section, we present some basic results on linear codes over the ring $R=\F_{q}+v\F_{q}+v^{2}\F_{q}$,
where $v^{3}=v$ and $q$ is odd \cite{gao,sol}.
The ring $R$ is equivalent to the ring $\frac{\F_{q}[v]}  {\langle v^3-v \rangle}$.
This shows that $R$ is a finite commutative, principal ring with the following non-trivial maximal ideals
\[
\langle v\rangle ,\, \langle 1-v \rangle,\, \langle 1+v \rangle.
\]
Hence by the Chinese Remainder Theorem we have
\begin{equation}
\label{eq:decom2}
R = R/\langle v\rangle \oplus R/\langle 1-v\rangle \oplus R/\langle1+ v\rangle.
\end{equation}
It is convenient to write the decomposition given in (\ref{eq:decom2}) using orthogonal idempotents in
$R$ which is given by
\begin{equation}
\label{decom:3}
R = \eta_1 R \oplus \eta_2 R \oplus \eta_3 R = \eta_1 \F_q \oplus \eta_2  \F_q  \oplus \eta_3  \F_q,
\end{equation}
where $\eta_1= 1-v^2$, $\eta_2=\frac{v+v^2}{2}$, and $\eta_3= \frac{v^2-v}{2}$.
Each element $r$ of $R$ can be expressed uniquely as $r=a_{0}+va_{1}+v^{2}a_{2}$, where $a_{i}\in \F_{q}, i=0, 1, 2$.

A linear code $C$ of length $n$ over $R$ is an $R$-submodule of $R^{n}$.
An element of $C$ is called a codeword of $C$.
A generator matrix of $C$ is a matrix whose rows generate $C$.
The Hamming weight $w_{H}(c)$ of a codeword $c$ is the number of nonzero components in $c$.
The Euclidean inner product is
\[
x \cdot y=x_{0}y_{0}+x_{1}y_{1}+\ldots +x_{n-1}y_{n-1},
\]
where $ x,y \in  R^{n}$.
The dual code $C^{\perp}$ of $C$ with respect to the Euclidean inner product is defined as
\[
C^{\perp}=\lbrace x\in R^{n};\,  x\cdot y=0, \forall y \in C\rbrace .
\]
A code $C$ is called self-dual if $C=C^{\perp}$ and $C$ is called self-orthogonal if $C\subseteq C^\perp$.

For a linear code $C$ of length $n$ over $R$, define
\[
\begin{array}{ccl}
C_{1} &=&\left\{ a\in \F_{q}^{n}; \, \exists b,c\in \F_{q}^{n}; \eta_1a+ \eta_2b+ \eta_3 c\in C\right\},\\
C_{2} &=&\left\{ b\in \F_{q}^{n}; \, \exists a,c\in \F_{q}^{n}; \eta_1a+ \eta_2b+ \eta_3 c\in C\right\}, \\
C_{3} &=&\left\{ c\in \F_{q}^{n}; \, \exists a,b\in \F_{q}^{n}; \eta_1a+ \eta_2b+ \eta_3 c\in C\right\}.
\end{array}
\]
It is clear that $ C_{1}$, $ C_{2}$ and $ C_{3}$ are linear codes of length $n$ over $\F_{q}$.
A direct consequence of the ring decomposition of $R$ in (\ref{decom:3}) is that a linear code $C$ over $R$
can be uniquely expressed as
\begin{equation}
\label{eq:sum}
C=\eta_1C_1 \oplus  \eta_2 C_2 \oplus\eta_3C_3.
\end{equation}
Moreover, from (\ref{eq:sum}) and the definition of the dual code we have that
\begin{equation}
\label{eq:sum1}
C^{\bot}=\eta_1C_1^{\bot} \oplus  \eta_2 C_2^{\bot} \oplus\eta_3C_3^{\bot}.
\end{equation}
Further, $C$ is self-dual if and only if $C_1$, $C_2$ and $C_3$ are self-dual over $\F_q$.
If $G_{1}$, $G_{2}$ and $G_{3}$ are generator matrices of $C_{1}$, $C_{2}$ and $C_{3}$, respectively, then
 \begin{equation}
 \label{generator/matrix1}
G=\left[
\begin{array}{c}
\eta_1 G_{1} \\
\eta_2 G_{2} \\
\eta_3 G_{3}
\end{array}
\right],
\end{equation}
is a generator matrix of $C$.
Often when working with codes over rings, an image to the underlying field is employed.
For the ring $R$ considered in this paper, a Gray map is defined as follows.
\begin{defi}
The Gray map $\Psi$ from $R^{n}$ to $\F^{3n}_{q} $ is defined by
\[
\begin{array}{c}
\Psi :R ^{n} \rightarrow \F^{3n}_{q}\\
(r_{0},r_{1}, \ldots,r_{n-1}) \mapsto (a_{0}, a_{0}+b_{0}+ c_{0},a_{0}- b_{0}+c_{0},\ldots,a_{n-1}+b_{n-1}+ c_{n-1},a_{n-1}-b_{n-1}+ c_{n-1}),
\end{array}
\]
where $r_{i}=a_{i}+vb_{i}+v^{2}c_{i}, i=0,1,\ldots,n-1$.
\end{defi}
For
$r=a+vb+v^{2}c$ in $R$, the Lee weight of $r$ is defined as
\[
w_{L}(a+vb+v^{2}c) =w_{H}(a, a +b +c, a- b +c),
\]
where $w_{H}$ denotes the Hamming weight of $v$ over $\F_{q}$.
Let $d_H$ denote the minimum Hamming distance of a code $C$.
For a codeword $c=( c_{0}, c_{1},\ldots, c_{n-1})$,
the Lee weight is defined as $w_{L}(c) = \sum \limits_{i=0}^{n-1}w_{L}( c_{i})$
and the Lee distance between codewords $c$ and $c'$ is defined as $d_{L}( c,c')=w_{L}( c-c')$.
The minimum Lee distance of a code $C$ is then $d_L(C) = \min d_L(c,c')$, $c \neq  c'$, $ \forall  c,  c'\in C$.

\begin{defi}
A linear code $C$ of length $n$ over $R$ and minimum Lee distance $d_L$ is called an $[n,|C|,d_L]_R$ code. Further if it is with minimum Hamming distance $d_H$, then it is denoted $[n,|C|,d_H]_R$
If $C$ has minimum Lee distance $d_L$ and is a free $R$-submodule that is isomorphic as a module to $R^k$,
then the integer $k$ is called the rank of $C$ and the code is denoted as $[n,k,d_L]_R$.
\end{defi}
\begin{prop}
\cite{sol}
\label{prop:1}
Let $ C $ be an $[n,\vert C \vert,d_{L}]_R$ code.
Then $\Psi(C)$ is a $[3n,k,d=d_L]$ linear code over $\F_{q}$.
Further, if $ C^{\perp} $ is the dual of $ C $, then $ \Psi\left(C \right) ^{\perp}=\Psi\left(C ^{\perp}\right) $.
\end{prop}

\begin{rem}
\label{rem:1}
If there exists an $[n,k,d_H]$ code $C$ over $\F_{q}$, then there exists a $[n,k,d_H]_R$ code
$\mathcal{C}= \eta_1 C \oplus  \eta_2 C \oplus\eta_3 C.$
\end{rem}
\begin{lem} \cite{sol}
\label{lem:1}
If $C$ is a linear code of length $n$ over $R$ with generator matrix $G$, then
\begin{equation}
\label{generator/matrix2}
\Psi(G)=\left[
\begin{array}{c}
\Psi(\eta_1 G_{1}) \\
\Psi(\eta_2 G_{2}) \\
\Psi(\eta_3 G_{3})
\end{array}
\right]=\left[
\begin{array}{ccc}
G_{1}&0&0 \\
0&G_{2}&0 \\
0&0& G_{3}
\end{array}
\right],
\end{equation}
and $d_H(\Psi(C))=\min \{ d_H( C_1), d_H(  C_2), d_H( C_3)\}$.
\end{lem}

We now give some useful results on cyclic codes over $R$.
A code $C$ is said to be cyclic if it satisfies
\[
\left( c_{n-1},c_{0},\ldots,c_{n-2}\right) \in C,\text{ whenever }\left(c_{0},c_{1},\ldots,c_{n-1}\right) \in C.
\]
It is well known that cyclic codes of length $n$ over $R$ can be considered as
ideals in the quotient ring $\frac{R[ x]}{ \langle x^{n}-1 \rangle}$
via the following $R$-module isomorphism
\[
\begin{array}{ccc}
R^{n} &\rightarrow &R\left[ x\right] \left/ \left\langle
x^{n}-1\right\rangle \right. \\
\left( c_{0},c_{1}, \ldots,c_{n-1}\right) &\mapsto
&c_{0}+c_{1}x+\ldots+c_{n-1}x^{n-1}
\end{array}
\]
\begin{defi}
The reciprocal of the polynomial $h(x)=h_{0}+ h_{1}x+\ldots+h_{k}x^{k}$ is defined as
\[
\begin{array}{ccc}
h^{\ast}(x)=x^{deg(h(x))} h(x^{-1}).
\end{array}
\]
If $h(x)=h^{\ast}(x)$, then the polynomial $h(x)$ is called self-reciprocal.
\end{defi}

\begin{prop}\cite{gao}
\label{thm:2}
Let $C$ be a cyclic code of length $n$ over $R$.
Then there exist polynomials $f_{i}(x)$ which are divisors of $x^{n}-1$ in $\F_{q}[x]$ such that
$C=\left\langle \eta_{1}f_{1}(x),\eta_{2} f_{2}(x),\eta_{3} f_{3}(x)\right\rangle$ and
$\vert C \vert=q^{3n-deg(f_{1}(x)+f_{2}(x)+f_{3}(x))}$.
Further
\[
C^{\bot}=\langle \eta_1 h^*_{1}(x), \eta_2 h^*_{2}(x),\eta_3 h^*_{3}\rangle,
\]
where $h_{i}(x) \in \F_{q}[x]$ such that $ x^{n}-1=f_{1}( x)h_{1}( x) =f_{2}( x)h_{2}( x)=f_{3}( x)h_{3}( x) $.
\end{prop}

\begin{prop}
There is no cyclic self-dual cyclic code of length $n$ over $R$.
\end{prop}
\pf
We know that $C=\eta_{1}C_{1}\oplus \eta_{2} C_{2}\oplus \eta_{3} C_{3}$.
From \cite[Theorem 1]{Jia1} we have that $C_{1}$, $C_{2}$ and $C_{3}$ are self-dual cyclic codes over $\F_{q}$
if and only if $q$ is a power of $2$ and $n$ is even. Since we assumed that $q$ is odd, the result then follows.
\qed

\section{LCD Codes over $R$}

A linear codes with complementary dual (LCD) code is defined as a linear code $C$ whose dual code
$C^{\bot}$ satisfies
\[
C \cap C^{\bot }=\lbrace0\rbrace.
\]
LCD codes have been shown to provide an optimum linear coding solution \cite{massey}.

For LCD codes over $R$, we have the following result.
\begin{thm}\label{thm:4}
A code $C=\eta_1 C_{1}\oplus \eta_2 C_{2}\oplus \eta_3 C_{3}$ of length $n$ over $R$ is an LCD code
if and only if $C_{1}$, $C_{2}$ and $C_{3}$ are LCD codes over $\F_q$.
\end{thm}
\pf
A linear code $C=\eta_1 C_{1}\oplus \eta_2 C_{2}\oplus \eta_3 C_{3}$ has dual code
$C^{\bot}=\eta_{1}C_{1}\oplus\eta_{2} C^{\bot}_{2}\oplus \eta_{3} C^{\bot}_{3}$.
We have that $ C\cap C ^{\bot}=\eta_{1}(C_{1}\cap C_{1} ^{\bot}) \oplus \eta_{2}(C_{2}\cap C_{2} ^{\bot}) \oplus \eta_{3}(C_{3}\cap C_{3} ^{\perp})$.
Due to the direct sum we have  $C \cap C^{\bot}=  \{0\}$ if and only if
$ C_{i}\cap C_{i} ^{\bot}= \{0\}$, for $ i=1,2,3 $.
Thus $C$ is an LCD code.
\qed

\begin{thm}
\label{thm:3}
If $C$ is an LCD code over $\F_q$, then $\mathcal{C}=\eta_{1} C \oplus \eta_{2} C \oplus \eta_{3}C $ is an LCD code over $R$.
If $\mathcal{C}$ is an LCD code of length $n$ over $R$, then $\Psi(\mathcal{C})$ is an LCD code of length $3n$ over $\F_q$.
\end{thm}
\pf
The first part is deduced from Theorem \ref{thm:4}.
From Proposition \ref{prop:1}, we have that $\Psi\left(C \right) ^{\bot}=\Psi(C ^{\bot})$.
Since $ \Psi $ is bijective and $\mathcal{C} \cap \mathcal{C}^{\bot }=\lbrace0\rbrace$, the result follows.
\qed

Next we give a necessary and sufficient condition on the existence of LCD codes over $R$.
First we require the following result due to Massey \cite{massey}.
\begin{prop}
\label{prop:Massey}
If $G$ is a generator matrix for an $[n,k]$ linear code $C$ over $\F_q$, then $C$ is an
LCD code if and only if the $k \times k$ matrix $GG^t$ is nonsingular.
\end{prop}
\begin{thm}
\label{lem:2}
If $G$ is a generator matrix for a linear code $C$ over $R$, then C is an
LCD code if and only if $GG^t$ is nonsingular.
\end{thm}
\pf
The generator matrix of $C$ can be expressed in canonical form as
\begin{equation}
 \label{generator/matrix}
G=\left[
\begin{array}{c}
\eta_1 G_{1} \\
\eta_2 G_{2} \\
\eta_3 G_{3}
\end{array}
\right].
\end{equation}
Since the $\eta_i$ are orthogonal idempotents, a simple calculation gives
\begin{equation}
\label{generator/matrixtrans1}
GG^t=\left[
\begin{array}{ccc}
\eta_1 G_{1} G_{1}^t&0&0 \\
0&\eta_2 G_{2}G_{2}^t&0 \\
0&0&\eta_3 G_{3}G_{3}^t.
\end{array}
\right].
\end{equation}
From Proposition \ref{prop:Massey} a necessary and sufficient condition for a code over $\F_q$ with generator matrix $G_i$ to be LCD
is that $G_iG_i^t$ be non singular.
Hence the proof follows from the generator matrix given in (\ref{generator/matrix}).
\qed

We now give conditions on the existence of cyclic LCD codes over $R$ using the generator polynomial.
This is an extension of the following result due to Massey \cite{massey}.

\begin{lem}
\label{massey:corps}
Let $C$ be a cyclic code over $\F_q$ generated by $f(x)$.
Then $C$ is LCD if and only if $f(x)$ is self-reciprocal.
\end{lem}
\begin{thm}
A cyclic code $C= \langle \eta_{1}f_{1}(x),\eta_{2} f_{2}(x),\eta_{3} f_{3}(x) \rangle$,
is an LCD code over $R$ if and only if for all $1\le i \le 3$, $f_i(x)$ is a self-reciprocal polynomial.
\end{thm}
\pf
The result follows from Proposition \ref{thm:2} and  Lemma \ref{massey:corps}.
\qed

\subsection{Existence of LCD Codes over $R$}
In this section, we show that LCD codes are an abundant class of codes over $R$.

\subsection{LCD Codes from Weighing Matrices}
In \cite{soleLCD}, the authors constructed LCD codes from orthogonal matrices and
left the existence of LCD codes from other classes of combinatorial objects as an open problem.
Thus, in this section we construct LCD codes over $\F_q$ and $R$ from weighing matrices.
We start with the following definition.
\begin{defi}
 \label{h}
A weighing matrix $W_{n, k}$ of order $n$ and weight $k$ is an $n \times n$
$(0, 1,-1)$-matrix such that $W W^t = kI_n$, $k \le n$.
A weighing matrix $W_{n,n}$, respectively $W_{n,n-1}$, is called a
Hadamard matrix, respectively conference matrix.
A matrix $W$ is symmetric if $W = W^t$, and $W$ is skew-symmetric (or skew) if $W =-W^t$.
\end{defi}
Tables of weighing matrices are given in \cite{CRC}.
Weighing matrices have been used to construct self-dual codes \cite{A-G}.
The following results show that it is also possible to construct LCD codes from weighing matrices.
\begin{prop}
\label{prop:weight}
Let $W_{n, k}$ be a weighing matrix of order $n$ and
weight $k$. We have the following results.
\begin{itemize}
\item[(i)] Let $\alpha$ be a nonzero element of $\F_q$ such that
$\alpha^2 + k \neq 0 \bmod q$.
Then the matrix
\begin{equation}
G=[\alpha I_{n} \;|\; W_{n, k}]
\end{equation}
generates an LCD $[2n, n]$ code over $\F_q$.
\item[(ii)] Let $W_{n, k}$ be a skew weighing matrix  of order $n$,
and $\alpha$ and $\beta$ nonzero elements of $\F_q$ such that $\alpha^2 +  \beta^2 +k\neq  0 \bmod q$.
Then the matrix
\begin{equation}
G=[\alpha I_{n}  \;|\; \beta I_{n} + W_{n, k}]
\end{equation}
generates a $[2n, n]$ LCD code over $\F_q$.
\end{itemize}
\end{prop}
\pf
The result follows from  Definition \ref{h} and Proposition \ref{prop:Massey}.
\qed
From Remark \ref{rem:1} and Proposition \ref{prop:weight} we have the following result.
\begin{cor}
Under the condition of Proposition \ref{prop:weight}, the matrix
\begin{equation}
\label{genmatrix}
\mathcal{G}=\left[
\begin{array}{c}
\eta_1 G \\
\eta_2 G \\
\eta_3 G
\end{array}
\right],
\end{equation}
is the generator matrix of a $[2n,n]$ LCD code over $R$.
\end{cor}

\begin{ex}
Let $q=3$, $n=6$, and $\alpha =2$ so that $\alpha^2 + 4 \neq 0 \bmod 3$.
Then for the weighing matrix given by
\[
W_{6,4}=\left(
\begin{array}{cccccccccc}
0&1&1&1&1&0\\
-1&0&0&1&-1&1\\
-1&0&0&-1&1&1\\
-1&-1&1&0&0&-1\\
-1&1&-1&0&0&-1\\
0&-1&-1&1&1&0
\end{array}
\right),
\]
$G=[2I_6 \;|\; W_{6,4}]$ generates a $[12,6,5]$ LCD code over $ \F_{3}$.
\end{ex}
Next we show that if $q$ is odd there always exists a suitable matrix to construct an LCD code.
\begin{thm}
\label{thm:lidl}
\cite[Theorem 7.32]{lidl}
Assume $q \equiv 3 \bmod 4$, $\eta$ be the quadratic character of $\F_{q}$ and $b_{ij}=\eta(j-i) $ for $ 1 \leqslant i $, $ j \leqslant q $, $ i\neq j $.
Then we have a Hadamard matrix given by
\begin{equation}\label{had:matrix}
H=\left[
\begin{array}{cccccc}
1&1&1&1&\ldots &1 \\
1&-1&b_{12}&b_{13}&\ldots&b_{1q} \\
1&b_{21}&-1&b_{23}&\ldots&b_{2q}\\
\vdots&\vdots&\vdots&\vdots&&\vdots\\
1& b_{q1}& b_{q2}& b_{q3}&\ldots&-1
\end{array}
\right].
\end{equation}
\end{thm}
\begin{cor}
\label{cor:Had}
For all nonzero $\alpha \in \F_q$ such that $q \equiv 3 \bmod 4$, the code generated by
 \begin{equation}
 \label{mat:Hada1}
  G=\left[\alpha I_{q+1}  \;|\; H \right],
 \end{equation}
where $H$ is the Hadamard matrix of order $q+1$ given in (\ref{had:matrix}),
is an LCD code over $\F_{q}$ of length $2(q+1)$.
\end{cor}
\pf
If $q \equiv 3 \bmod 4$, then from \cite[Lemma 3.3]{GG} $\alpha^2 +1 = 0$ has no solution in $\F_q$.
The result then follows from Proposition \ref{prop:weight} and Theorem \ref{thm:lidl}.
\qed
\begin{ex}
From Corollary \ref{cor:Had}, the following Hadamard matrix over $\F_3$
\[
H_{4}=\left[
\begin{array}{ccccc}
1&1&1&1\\
1&-1&-1&1\\
1&1&-1&-1\\
1&-1&1&-1
\end{array}
\right]
\]
gives an $[8,4,4]$ LCD code over $\F_3$.
According to \cite{table} this is an optimal code.
\end{ex}
\begin{thm}\cite[p. 56]{macwilliams}
\label{thm:MacWilliams}
Assume $q \equiv 1 \bmod 4$, $\eta$ be the quadratic character of $\F_{q}$ and $b_{ij}=\eta(j-i)$ for
$1 \leqslant i $, $ j \leqslant q $.
Then we have a symmetric conference matrix given by
\begin{equation}\label{conf:matrix}
Q=\left[
\begin{array}{cccccc}
0&1&1&1&\ldots &1 \\
1&b_{11}&b_{12}&b_{13}&\ldots&b_{1q} \\
\vdots&\vdots&\vdots&\vdots&&\vdots\\
1& b_{q1}& b_{q2}& b_{q3}&\ldots& b_{qq}
\end{array}
\right].
\end{equation}
\end{thm}
\begin{cor}
\label{cor:paley}
For all nonzero $\alpha \in \F_q$ such that $q \equiv 1 \bmod 4$,
the code generated by
 \begin{equation}
 \label{mat:Hada2}
  G=\left[ \alpha I_{q+1}  \;|\; Q \right],
 \end{equation}
where $Q$ is the conference matrix of order $q+1$ given in (\ref{conf:matrix}),
is an LCD code over $\F_{q}$ of length $2(q+1)$.
\end{cor}
\pf
The result follows from  Theorem \ref{thm:MacWilliams} and Proposition \ref{prop:Massey}.
\qed

In \cite{A-G} the authors constructed self-dual codes from conference matrices.
We note that if $\alpha^2+k=0$ has a solution, the matrix $[\alpha I_n  \;|\; W_{n,k}]$ generates a self-dual code over $\F_q$.
Then if there exists $\alpha' \neq 0$ such that $\alpha'^2+k\neq 0$,
from Proposition \ref{prop:weight} $[\alpha' I_n  \;|\; W_{n,k}]$ generates an LCD code with the same parameters as the self-dual code.
This result also holds for the minimum distance of LCD codes generated by $G=[\alpha' I_n  \;|\; \beta I_n + W_{n, k}]$.
It is easy to verify that whenever we have a skew matrix $W_{n,k}$, we can construct a skew matrix $W_{2n,2k+1}$, where
 \begin{equation}
\label{generator/matrixtrans2}
W_{2n,2k+1}=\left[
\begin{array}{cc}
W_{n,k}& -W_{n,k}-I \\
W_{n,k}+I& W_{n,k}
\end{array}
\right].
\end{equation}

The above results were used to construct the LCD codes over $\F_p$, $p$ prime, $3< p\le 23$, given in Tables 1-4.
It is worth noting that for many parameters a self-dual code cannot be constructed from weighing matrices,
whereas for the same parameters (except for the case $p=3$), it was always possible to construct LCD codes.

\begin{table}
\begin{center}
\caption{LCD codes from Proposition~\ref{prop:weight} using conference matrices with $N=8$, $12$ and $16$.}
\label{table5}
\begin{tabular}{|c|c|c|c|c|c|c|c|c|c|c|c|c|}
\hline
$p$ & $\alpha $ & $\beta$ & $ d$ &  $ \alpha $  & $ d$ & $\alpha $  & $ \beta $ & $ d$  \\
\hline
$5$ & $2$ & $1$ & $4$  & $1$ & $6$ & $2 $ & $1 $ & $7 $  \\
$7$ & $1$ & $3$ & $5$  & $1$ & $6$ & $ 2$ & $1 $ & $7 $ \\
$11$ & $1$ & $2$ & $5$ & $1$ & $6$ & $ 1$ & $0 $ & $ 7$ \\
$13$ & $2$ & $3$ & $5$ & $1$ & $6$ & $ 1$ & $ 6$ & $7 $  \\
$17$ & $2$ & $8$ & $5$ & $1$ & $6$ & $ 2$ & $3 $ & $ 7$  \\
$19$ & $1$ & $8$ & $5$ & $1$ & $6$ & $ 2$ & $ 7$ & $ 7$ \\
$23$ & $3$ & $4$ & $5$ & $1$ & $6$ & $ 3$ & $0 $ & $ 7$ \\
\hline
\end{tabular}
\end{center}
\end{table}
\begin{table}
\begin{center}
\caption{LCD codes from Proposition~\ref{prop:weight} using conference matrices with $N=20$, $24$ and $28$.}
\label{table2}
\begin{tabular}{|c|c|c|c|c|c|c|c|c|c|c|c|}
\hline
$p$ & $\alpha $ & $ d$  &$ \alpha $ & $\beta $ & $ d$  &$\alpha $ & $ d$  \\
\hline
$5$ & $2$ & $8$ & $1$ & $0$ & $9$ & $ 1 $&$ 10 $ \\
$7$ & $1$ & $8$ & $2$ & $3$ & $9$ & $ 2 $&$ 10 $\\
$11$ & $1$ & $8$ & $1$ & $0$ & $9$ &  $ 1 $& $10$\\
$13$ & $1$ & $8$ & $5$ & $5$ & $9$ &  $ 1 $&$ 10 $\\
$17$ &$1$ & $8$ & $1$ & $6$ & $9$ & $1  $& $10  $\\
$19$ &$1$ & $8$ & $2$ & $8$ & $9$ & $1  $&$10  $\\
$23$ &$1$ & $8$ &$1$ & $0$ & $9$ & $1 $&$10 $\\
\hline
\end{tabular}
\end{center}
\end{table}
\begin{table}
\begin{center}
\caption{LCD codes from Proposition~\ref{prop:weight} using conference matrices with $N=32$, $36$ and $40$.}
\label{table3}
\begin{tabular}{|c|c|c|c|c|c|c|c|c|c|c|c|}
\hline
$p$ & $\alpha$ & $ \beta $ & $ d$  & $ \alpha $ & $ d$ & $\alpha $ & $ \beta $& $ d$  \\
\hline
$5$ & $2$ & $2$ & $10$ & $1$ & $12 $ &$2$ & $0$ & $13$ \\
$7$ & $ 1$ & $3$ & $11$ & $1$ & $12$ &$1$ & $0$ & $13$ \\
$11$ & $1$ & $5$ & $11$ & $1$ & $12$ & $1$ & $0$ & $13$ \\
$13$ & $2$ & $6$ & $11$ &   $1$ & $12$ & $1$ & $4$ & $13$ \\
$17$ & $1$ & $0$ & $11$ &  $1$ & $12$ & $1$ & $0$ & $13$ \\
$19$ & $1$ & $0$ & $11$ &  $1$ & $12$ & $1 $ & $0 $ & $13$ \\
$23$ & $1$ & $2$ & $11$ &  $1$ & $12$ & $1$ & $0$ & $13$ \\
\hline
\end{tabular}
\end{center}
\end{table}

\begin{table}
\begin{center}
\caption{LCD codes from Proposition~\ref{prop:weight} using the skew matrix $W_{14,9}$.}
\label{table4}
\begin{tabular}{|c|c|c|c|}
\hline
$p$ & $\alpha$ & $ \beta $ & $ d$   \\
\hline
$5$ & $2$ & $0$ & $8$  \\
$7$ & $ 2$ & $2$ & $10$  \\
$11$ & $1$ & $3$ & $10$ \\
$13$ & $2$ & $4$ & $11$ \\
$17$ & $1$ & $2$ & $11$ \\
$19$ & $2$ & $3$ & $11$  \\
$23$ & $2$ & $6$ & $11$ \\
\hline
\end{tabular}
\end{center}
\end{table}

\subsection{General Construction of LCD Codes}
We start with the following lemma.
\begin{lem}
\label{lem:LCD}
If $R= \F_q+v\F_q+v^2\F_q$ with $q=p^r$ a power of an odd prime, then the following hold:
\begin{itemize}
\item[(i)] there exists $\alpha \in R$ such that $\alpha^2+1=0$ if  $q\equiv 1 \bmod 4$,
\item[(ii)] there exist $\alpha, \beta \in R$ such that $\alpha^2 +\beta^2+1=0$ if  $q\equiv 3 \bmod 4$, and
\item[(iii)] for every $q$ there exist $\alpha, \beta,\gamma, \delta \in R$ such that $\alpha^2+ \beta^2+\gamma^2+\delta^2 =0 $ in $R$.
\end{itemize}
\end{lem}
\pf
It is easy to show that if there exist solutions over $\F_q$ for cases (i)-(iii),
then these solutions also hold over $R$ since $\F_q$ is a subring of $R$.
Hence we only need to show that these solutions exist over $\F_q$.
From \cite[Lemma 3.3]{GG}, if $q\equiv 1 \bmod 4$ then $-1$ is a square in $\F_q$, which proves (i).
From \cite[p. 281]{rosen}, if  $q\equiv 3 \bmod 4$ then there exist $\alpha, \beta \in \F_q$ such that $\alpha^2+ \beta^2+1=0$, which proves (ii).
From \cite[Theorem 370]{hardy}, we have that every prime is the sum of four squares. Since $q=p^r$ and $p=0$ in $R$ this proves (iii).
\qed

The next result shows that it is always possible to construct LCD codes over $R$.
\begin{thm}
\label{prop:LC}
If $P$ is the generator matrix of a self-dual code over $R$, then the generator matrix $G = [I \;|\; P]$ generates an LCD code over $R$.
If $G=[I \;|\; P]$ is the generator matrix of a linear code over $R$, then the following hold.
\begin{itemize}
\item[(i)] If $q\equiv 1 \bmod 4$ and $\alpha^2+1=0$, then the code over $R$ with generator matrix $G'= [I  \;|\; P  \;|\; \alpha P]$ generates an LCD code over $R$.
\item[(ii)] If $q\equiv 3 \bmod 4$ and $\alpha, \beta \in \F_q$ such that $\alpha^2 +\beta^2+1=0$, then
$G'= [I  \;|\; P  \;|\; \alpha P   \;|\; \beta P]$ generates an LCD code over $R$.
\item[(iii)] If $R= \F_q+v\F_q+v^2\F_q$ with $q=p^r$, then  $G'= [I  \;|\; P  \;|\; \alpha P   \;|\; \beta P  \;|\; \delta P  \;|\; \gamma P]$ such that $\alpha^2+ \beta^2+\gamma^2+\delta^2 =p$
generates an LCD code over $R$.
\end{itemize}
\end{thm}
\pf
Part (i) is just a verification. The other parts follow from Lemma \ref{lem:LCD}.
\qed

\section{Construction of Formally Self-Dual Codes over $R$}
Recall that a code $C$ is called  formally self-dual if $C$ and $C^{\bot }$ have the same weight enumerator.
Codes which are equivalent to their dual are called isodual codes,
and isodual codes are also formally self-dual.
In this section, we present three constructions of formally self-dual codes over $R$.
First, we give the following result which links formally self-dual codes over $R$ to formally self-dual codes over $\F_q$.

\begin{thm}
If $C$ is a formally self-dual code over $R$, then the image under the corresponding Gray map is a formally self-dual code.
\end{thm}
\pf
The result follows from Theorem \ref{thm:2} and the fact that the Gray map is an isometry.
\qed

An $ n \times n $ square matrix $ M$ is called $\lambda$-circulant of order $n$ if it has the following form
\[
M=\left[
\begin{array}{cccccc}
 M_{11} & M_{12} & M_{13}& \ldots & M_{1n} \\
\lambda M_{1n} & M_{11} & M_{12}& \ldots & M_{1n-1} \\
\lambda M_{1n-1} & \lambda M_{1n} & M_{11}& \ldots & M_{1n-2} \\
\vdots & \vdots & \vdots & \vdots \\
\lambda M_{12} & \lambda M_{13} & \lambda M_{14} & \ldots & M_{11}
\end {array}
\right].
\]
If $\lambda=1$ this matrix is circulant and there is a vast literature on double circulant and bordered
double circulant self-dual codes \cite{Batoul}.

The proof of the next theorem is the same as that for \cite[ Theorem 6.1]{Batoul} given over the ring $\F_q+v\F_q$.
It is given here for completeness.
\begin{thm}\label{thm:6}
Let $M$ be a $\lambda $-circulant matrix over $R$ of order $n$.
Then the code generated by $G=[ I_{n}  \;|\;  M]$ is a formally self-dual code over $ R $.
\end{thm}
\pf
Let $C$ be the code generated by $G=[ I_{n} \;|\; M]$ and $C'$ the code generated by $G'=[ M^{t}  \;|\; -I_{n}]$.
It is easy to verify that the codes $C$ and $ C'$ are orthogonal codes, and since they both have rank $n$, $ C'=C^{\bot} $.
Let $C''$ be the code generated by $G''=[ M^{t}  \;|\; I_{n}]$.
Since $ wt(a) = wt (-a)$ for any $a$ in $R$, the codes $ C' $  and $C'' $ have the same weight enumerator.
In order to complete the proof, we show that $ C''$ is equivalent to $C$, therefore $C$ is formally self-dual.
Let $\sigma$ be the permutation
\[
\sigma = (1, n) (2, n-1)\ldots(k-1, n-k + 2) (k, n-k + 1),
\]
where $k =n/2$.
If $M'$ is the matrix obtained by applying $\sigma$ on the rows of $ M$ and $ M''$ is the matrix obtained by applying $\sigma$ on the columns of $M'$,
then $M'' = M^{t}$.
Hence, $M$ and $M^{t}$  are equivalent.
Similarly, by applying a suitable column permutation we obtain that $G$ and $G''$ are equivalent.
Thus, $C$ and $C''$ are equivalent and therefore $C$ is formally self-dual.
\qed

\begin{ex}
 Let $q=3$, $n=4$, $ \lambda=1+v $ and $ M $ be the following $ \lambda $-circulant matrix
\[
M=\left[
\begin{array}{cccccccccc}
2v+2v^{2}&2+v+v^{2}&1+2v&2\\
2+2v&2v+2v^{2}&2+2v+2v^{2}&1+2v\\
1+2v^{2}&2+2v&2v+2v^{2}&2+2v+2v^{2}\\
2+v^{2}&1+2v^{2}&2+2v&2v+2v^{2}\\
\end{array}
\right].
\]
Then $G=[ I_{4}  \;|\; M ]$ generates a formally self-dual code of length $8$ over $ \F_{3}+v\F_{3} +v^{2}\F_{3}$.
The Gray image of this code is a $[24,12,9]$ formally self-dual code over $\F_3$. This is an optimal code.
\end{ex}

\begin{ex}
Let $q=5$, $n=3$, $ \lambda=2+v $ and $ M $ be the following $ \lambda $-circulant matrix
\[
M=\left[
\begin{array}{cccccccccc}
3v+2v^{2}&4v&3+2v\\
1+2v+2v^{2}&3v+2v^{2}&4v\\
3v+4v^{2}&1+2v+2v^{2}&3v+2v^{2}\\
\end{array}
\right].
\]
Then $ [ I_{3}  \;|\; M ] $ generates a formally self-dual code of length $6$ over $ \F_{5}+v\F_{5} +v^{2}\F_{5}$.
The Gray image of this code is an $\left[ 18,9,7\right]$ formally self-dual code over $\F_5$.
\end{ex}

\begin{thm}\label{thm:7}
Let $M$ be an $(n-1)\times (n-1)$  $ \lambda $-circulant matrix.
Then the code generated by
\[
G=\left[  I_{n}\left\vert
\begin{array}{cccc}
\alpha & \omega & \ldots & \omega \\
\omega &  &  &  \\
\vdots &  & M &  \\
\omega &  &  &
\end{array}%
\right. \right]
\]
where $\alpha, \omega \in R$, is a formally self-dual code over $R$.
\end{thm}
\pf
The proof is similar to that of Theorem \ref{thm:6}.
\qed
\begin{ex}
Let $q=3$, $\alpha=2+v+2v^{2}$, $\omega=2+2v$, $ \lambda=1+v^{2} $, $n=3$ and
\[
M=\left[
\begin{array}{cccccccccc}
2&1+v\\
1+2v+v^{2}&2\\
\end{array}
\right].
\]
Then
\[
G=\left[
\begin{array}{cccccccc}
1 & 0 & 0 & 2+v+2v^{2} & 2+2v& 2+2v  \\
0 & 1 & 0 & 2+2v & 2 & 1+v  \\
0 & 0 & 1 & 2+2v & 1+2v+v^{2}&2 \\

\end{array}
\right],
\]
generates a formally self-dual code of length $6$ over $ \F_{3}+v\F_{3} +v^{2}\F_{3}$.
The Gray image of this code is an $[18,9,6]$ formally self-dual code over $\F_3$.
This is an optimal code.
\end{ex}

Using a proof similar to that of Theorem \ref{thm:6}, we have the following construction of formally self-dual codes over $R$.
\begin{thm}
\label{thm:5}
Let $A$ be an $n\times n$ matrix over $R$ such that $A^{t}=A$.
Then the code generated by $G=\left[ I_{n}\,|\, A \right] $ is a formally self-dual code over $ R $ of length $2n$.
\end{thm}

\begin{ex}
Let $q=5$, $n=3$, and $A$ be the matrix
\[
A=\left[
\begin{array}{ccccccc}
4v+1&v+v^2&4+4v\\
v+v^2&4v&1+v\\
4+4v&1+v&1+v\\
\end{array}
\right].
\]
We have that $A=A^{t}$, so $[ I_{3} \;|\; A ]$ generates a formally self-dual code of length $6$ over $ \F_{5}+v\F_{5} +v^{2}\F_{5}$.
The Gray image of this code is an $[18,9,7]$ formally self-dual code over $\F_{5}$.
\end{ex}
\begin{ex}
Let $q=9$, $n=5$, and $A$ be the matrix
\[
A=\left[
\begin{array}{ccccccc}
0&v&8+v&1+8v+8v^{2}&8v+8v^{2}\\
v&8v+8v^{2}&8&1+v&1+v^{2}\\
8+v&8&8v^{2}&8+v+v^{2}&1+8v\\
1+8v+8v^{2}&1+v&8+v+v^{2}&1&v\\
8v+8v^{2}&1+v^{2}&1+8v&v&8\\
\end{array}
\right].
\]
We have that $A=A^{t}$, so $[ I_{5} \;|\; A ]$ generates a formally self-dual code of length $10$ over $ \F_{9}+v\F_{9} +v^{9}\F_{9}$.
The Gray image of this code is a $\left[ 30,15,12 \right]$ formally self-dual code over $\F_{9}$.
\end{ex}

\begin{prop}
Let $C_{1}$, $C_{2}$ and $C_{3}$ be linear codes of length $n$ over $\F_q $.
$C_{1}$, $C_{2}$ and $C_{3}$ are isodual codes if and only if $C=\eta_1C_{1}\oplus \eta_2 C_{2}\oplus  \eta_3 C_{3}$
is an isodual code of length $n$ over $R$.
\end{prop}
\pf
Let $\tau_1$, $\tau_2$, and $\tau_2$ be monomial permutations such that $\tau_1(C_1)= C_1^\bot$, $\tau_2(C_2)=C_2^\bot$ and $ \tau_3(C_3)=C_3^\bot$.
Then $\eta_1\tau_1(C_1)\oplus \eta_2\tau_2(C_2)\oplus \eta_3 \tau_3(C_3)=\eta_1C_1^\bot\oplus \eta_2C_2^\bot \oplus \eta_3C_3^\bot$.
Since $C^\bot =\eta_1C_1^\bot\oplus \eta_2 C_2^\bot \oplus \eta_3 C_3^\bot$, it follows that $C$ is equivalent to $C^\bot$.
\qed

\subsection{Construction of LCD Formally Self-Dual Codes over $R$}

LCD formally self-dual codes over $R$ are constructed in this subsection.

\begin{thm}
With the same assumptions as in Theorem \ref{thm:6},
the matrix $G=[I_{n} \vert M]$ generates an LCD formally self-dual code over $R$ if and only if $GG^t$ is nonsingular.
\end{thm}
\pf
The result follows from Theorem \ref{thm:6} and Theorem \ref{lem:2}.
\qed
\begin{ex}
Let $q=5$, $n=4$, $ \lambda=4v^{2} $ and $ M $ be the following $\lambda$-circulant matrix
\[
M=\left[
\begin{array}{ccccccc}
2v^{2} & 0 & v & 0 \\
0 & 2v^{2} & 0 & v \\
4v & 0 &2v^{2} &0\\
0& 4v& 0 & 2v^{2}\\
\end{array}
\right].
\]
It is easily determined that $GG^t$ is nonsingular.
Thus, $G=[ I_{4}\left\vert M\right] $ generates an LCD
formally self-dual code of length $8$ over $\F_{5}+v\F_{5}+v^{2}\F_{5}$.
\end{ex}
\begin{thm}
 With the same assumptions as in Theorem \ref{thm:7}, the code generated by
\[
G=\left[  I_{n}\left\vert
\begin{array}{cccc}
\alpha & \omega & \ldots & \omega \\
\omega &  &  &  \\
\vdots &  & M &  \\
\omega &  &  &
\end{array}%
\right. \right],
\]
is an LCD formally self-dual code over $R$ if and only if $GG^t$ is nonsingular.
\end{thm}

\begin{ex}
Let $q=3$, $n=3$, $\lambda=v^{2} $ and $ M $ be the following $ \lambda $-circulant matrix
\[
M=\left[
\begin{array}{ccc}
v& v  \\
v & v
\end{array}
\right].
\]
If $\alpha=v$ and $\omega=v^{2}$, it is easily determined that for the matrix
\[
G=\left[
\begin{array}{cccccccc}
1 & 0 & 0 & v & v^{2} & v^{2}   \\
0 & 1 & 0 &  v^{2}& v& v \\
0 & 0 & 1 & v^{2} & v & v\\

\end{array}
\right],
\]
$GG^t$ is nonsingular.
Then $G$ generates an LCD formally self-dual code of length $6$ over $\F_{3}+v\F_{3}+v^{2}\F_{3}$.
\end{ex}

\begin{thm}
\label{thm:8}
With the same assumptions as in Theorem \ref{thm:5},
the code generated by $G=[ I_{n} \vert A ] $ is an LCD formally self-dual code over $R$ if and only if $GG^t$ is nonsingular.
\end{thm}
\pf
From Theorem \ref{thm:5}, we know that $G$ generates a formally self-dual code.
To prove that $G$ generates an LCD codes we apply the conditions of Theorem \ref{lem:2} on the matrix $ G=[ I_{n} \vert  A]$.
\qed

\section{Bounds on LCD Codes}
Bounds on codes are important for the combinatorial properties of the codewords as it was shown in \cite{BGM}. In this section we will give some bounds on LCD codes over fields and the ring $R$. When dealing with codes over $\F_q$, $q$ may be odd or even. While, when we are over the ring $R$, $q$ must be odd as assumed in the begin of the paper.

We begin with the following bound which was given by Dougherty et al. \cite{soleLCD}
\[
LCD[n, k]_q := \max\{d; \text{ there exists an } [n, k, d]_q \text{ LCD code}\}.
\]
They also gave estimates and results for this bound.
For linear codes over $\F_q$, we have the following bound
\[
\mathcal{B}(n,k)_q :=  \max\{d; \text{ there exists an } [n, k, d]_q \text{ code}\}.
\]
The next result gives an upper bound on $d$ for an LCD code over $\F_q$
\begin{prop}
If an $[n,n/2,d]_q$ self-dual code $C$ exists, then $LCD[3n/2,n/2]_q \ge d$.
In particular, if $C$ is an extremal self-dual code over $\F_2$,
then $LCD[3n/2,n/2]_2 \ge \frac{4n}{24} +4$ if $n \neq 22  \bmod 24$,
and
$LCD[3n/2,n/2]_2 \ge \frac{4n}{24} +6$ if $n\equiv 22 \bmod 24$.
For $\F_3$ and $n \equiv 0 \bmod 4$, $LCD[3n/2,n/2]_3 \ge 3n/12 +3$.
\end{prop}
\pf
If there exists an $[n,n/2,d]_q$ self-dual code with generator matrix $P$,
then the matrix $G=[I \;|\; P]$ satisfies $GG^t=I$, and hence by Proposition \ref{prop:Massey} $G$ generates an LCD code with parameters $[3n/2, n/2, \ge d]$.
In the binary case, the bound given corresponds to the condition on extremal binary self-dual codes given in \cite[Chap. 9 ]{huffman03}.
In the ternary case, the bound corresponds to the condition on extremal ternary self-dual codes given in \cite[Chap. 9 ]{huffman03}.
\qed

The proof of the following result follows from Theorem \ref{prop:LC} and the Singleton bound.
\begin{prop}
\label{th:bound}
If there exists an $[n,k,d]_q$ code over $\F_q$, then
\[
\begin{array}{ccl}
d&\le & LCD[n+k, k]_q  \le \mathcal{B}(n+k,k)_q \le n+1 \mbox{ if } q=2^m,\\
d&\le & LCD[2n+k, k]_q  \le \mathcal{B}(2n+k,k)_q \le 2n+1 \mbox{ if } q\equiv 1 \bmod 4,\\
d&\le & LCD[3n+k, k]_q  \le \mathcal{B}(3n+k,k)_q \le 3n+1 \mbox{ if } q\equiv 3 \bmod 4,\\
d&\le & LCD[4n+k, k]_q  \le \mathcal{B}(4n+k,k)_q \le 4n +1 \mbox{ for all } q.
\end{array}
\]
\end{prop}

\begin{prop}
If $q$ is odd, then $LCD[q+1, q-2\mu]_q=B(q+1,q- 2\mu)_q= 2\mu+2$, for $1 \le \mu \le \frac{q-1}{2}$.

If $q$ is even, then $LCD[q+1, q-1-2\mu]_q=B(q+1,q-1- 2\mu)= 2\mu+3$, for $1 \le \mu \le \frac{q-1}{2}-1$.
\end{prop}
\pf
From \cite[Theorem 8]{grassl}, if $q+1-k$ is odd (this case correspond to $q$ and $k$ both even or odd),
then the cyclic code generated by the polynomial $g_1(x)= \prod_{i=- \mu}^\mu(x-\alpha ^{i})$ is a
$[ q+1,q-2\mu, 2\mu+2]_q$ maximum distance separable (MDS) cyclic code.
Since $g_1(x)$ is self-reciprocal, from Lemma \ref{massey:corps} the code is LCD.
Since the code is also MDS, the result follows.

If $q$ is even and $k$ is odd, then the polynomial $g_2(x)= \prod_{i=q/2- \mu}^{q/2}(x- \alpha ^{i})(x-\alpha ^{-i})$
generates a $[ q+1,q-1-2\mu, 2\mu+3]_q$ MDS cyclic code from \cite[Theorem 8]{grassl}.
Since $g_2(x)$ is self-reciprocal, the code is LCD by Lemma \ref{massey:corps}.
The result then follows since the code is MDS.
\qed

For codes over $R$, define the bound
\[
LCD[n, k]_R := \max\{d; \text{ there exists an} [n,k,d_{H}]_R\text{ free LCD code over }R\}.
\]
From Remark \ref{rem:1}, we have the following bound
\begin{equation}
\label{eq:fin}
LCD[n, k]_R \ge LCD[n, k]_q,
\end{equation}
which gives the following result.
\begin{cor}
All the lower bounds on $LCD[n, k]_q$ given in \cite{soleLCD} are also lower bounds on $LCD[n, k]_R$.
\end{cor}

\section{Conclusion and Open Problems}

In this paper, LCD codes and formally self-dual codes were considered over the ring $R=\F_{q}+v\F_{q}+v^{2}\F_{q}$, where $v^{3}=v$, for $q$ odd.
Conditions were given on the existence of LCD codes,
and constructions presented for LCD and formally self-dual codes over $R$.
Some of the results presented can easily be generalized to codes over some Frobenius rings such as those in \cite{waifi}.
As an extension of the results in \cite{soleLCD}, we gave bounds on LCD codes over $\F_q$,
and used these to obtain bound on LCD codes over $R$.
It was shown that lower bounds on LCD codes over $\F_q$ are also lower bounds on LCD codes over $R$.
LCD codes were constructed from weighing matrices.
It will be interesting to construct LCD codes from other combinatorial objects.
Further, it should be possible to obtain a linear programming bound for codes over $R$.
\begin{center}
\textbf{Acknowledgements}
\end{center}

The authors would like to thank reviewers as well as Professors R. K. Bandi, S. Dougherty, A. Kaya and I. Kotsireas for their useful comments which improved considerably the paper.
 This research is supported by National Natural Science Foundation of China (61672036), Technology Foundation for Selected Overseas Chinese Scholar, Ministry of Personnel of China (05015133), the Open Research Fund of National Mobile Communications Research Laboratory, Southeast University(2015D11) and Key projects of support program for outstanding young talents in Colleges and Universities (gxyqZD2016008).

\end{document}